\documentclass[conference]{IEEEtran}
\usepackage{cite}
\usepackage{amsmath,amssymb,amsfonts}
\usepackage{subcaption}
\usepackage{algorithm}
\usepackage{algorithmic}
\usepackage{graphicx}
\usepackage{textcomp}
\usepackage{xcolor}
\usepackage{url}

\newcommand{\at}{\makeatletter @\makeatother}

\def\BibTeX{{\rm B\kern-.05em{\sc i\kern-.025em b}\kern-.08em
    T\kern-.166em\lower.7ex\hbox{E}\kern-.125emX}}
\begin{document}

\title{
Dijkstra-Through-Time: Ahead of time hardware scheduling method for
deterministic workloads
}
\author{\IEEEauthorblockN{Vincent Tableau Roche}
\IEEEauthorblockA{\textit{Nokia Bell Labs} \\
Antwerp, Belgium \\
vincent.tableau\_roche.ext\at nokia.com}
\and
\IEEEauthorblockN{Purushotham Murugappa Velayuthan}
\IEEEauthorblockA{\textit{Nokia Bell Labs} \\
Antwerp, Belgium \\
purushotham.mv@nokia-bell-labs.com}
}

\maketitle
\begin{abstract}
Most of the previous works on data flow optimizations for Machine Learning hardware accelerators try to find algorithmic re-factorization such as loop-reordering and loop-tiling. However, the analysis and information they provide are still at very high level and one must further map them onto instructions that hardware can understand. This paper presents ``Dijkstra-Through-Time" (DTT), an ahead of time compute and memory scheduling-mapping algorithm for deterministic workloads. It provides a simple implementation and supports accelerators with complex NoC configurations, at the expense of a long compilation process. This initial paper illustrates a proof of concept implementation to merge scheduling and data cache coherence mechanisms to get more optimized data flows.
\end{abstract}
\section{Introduction}
In recent years there has been a Cambrian explosion of hardware accelerators
particularly related to area of  Machine Learning(ML)\cite{mlsurvey_2019}. One
of the driving factors for this is the wide and ever-increasing application domain of
machine learning, with each application imposing its unique size,
cost, latency, accuracy, and power performance requirements. Additionally,
the flexibility to support various or a select class of Neural Network(NN) models
together with the ability to efficiently perform training along with inference  are
considered desirable \cite{Sze2020}. Recent works such as nGraph\cite{ngraph},
Glow\cite{glow} and TVM\cite{tvm} aim to solve the issue of software compiler
support for custom NN accelerators, while tools such as ZigZag\cite{zigzag},
MAESTRO\cite{maestro} and TimeLoop\cite{timeloop} implement complex
hardware-aware data flow optimizations.

Most accelerators often rely on a multi-core architecture to perform
computations in parallel. In many of these multi-core
architectures, data movement follows a design philosophy similar to that of the
original Stanford DASH processor\cite{dash}. DASH uses a cache coherence
protocol that ensures the right copy of a page will always be used in the
hardware, even if very little determinism can be assumed from the processing
elements (called PEs henceforth). 

In practice however, one might notice that the behavior of the PEs is in fact
determined by the (known) program they execute. If the code sequence is deterministic 
(i.e. with no data dependent conditional codes present) one could try to read the
programs of the PEs and create an optimized data flow that guarantees cache
coherence without any classic hardware-implemented data consistency protocol.

A precedent for such a data flow can be found in the field of supply chain
management which defines Pulled and Pushed flow\cite{supplyChainFlows}. In a
\textbf{pulled flow}, the goods are only produced when ordered by a consummer,
which is often preferred in practice because it scales well with uncertainty.
On the other hand, in a \textbf{pushed flow} goods are produced before they are
ordered and stored until purchase. \textbf{Pushed flows} create determinism in
the supply chain, enabling economies of scale by producing or shipping the goods
in bulk.

In the context of NN accelerator design, a typical implementation would be with PEs 
requesting data explicitly and receiving pages of data to work on i.e. with a
\textbf{pulled flow}. In this paper, we investigate the \textbf{pushed flow}, where the
PEs receive the data without explicitly requesting it. The idea stems from the fact that most NN 
workloads are deterministic in nature without conditional statements. Hence, this determinism 
can be exploited to pre-schedule workloads across PEs in an efficient way.

In this paper we are presenting ``Dijkstra-Through-Time" (DTT) a python-based tool as a proof 
of feasibility of this idea. As the name indicates, DTT applies Dijkstra's algorithm 
on a snapshot of system states taken at every instant. As these are our first results, we present the methodology with some qualitative analysis and leave quantitative comparison with practical NN models as future work.

The rest of the paper is organized as follows: In the first section, 
we present a high-level intuition of the DTT method, taking the 
implementation into account. The second section provides examples of using the 
algorithm to clarify the way DTT works, while in the final section we are going
to present what a theoretical architecture leveraging the DTT algorithm would
look like.
\section{Algorithm Overview}
\subsection{Intuition of the Method}
\begin{figure}[tbp]
    \centering
    \scalebox{0.8}[0.6]{\includegraphics[width=\linewidth]{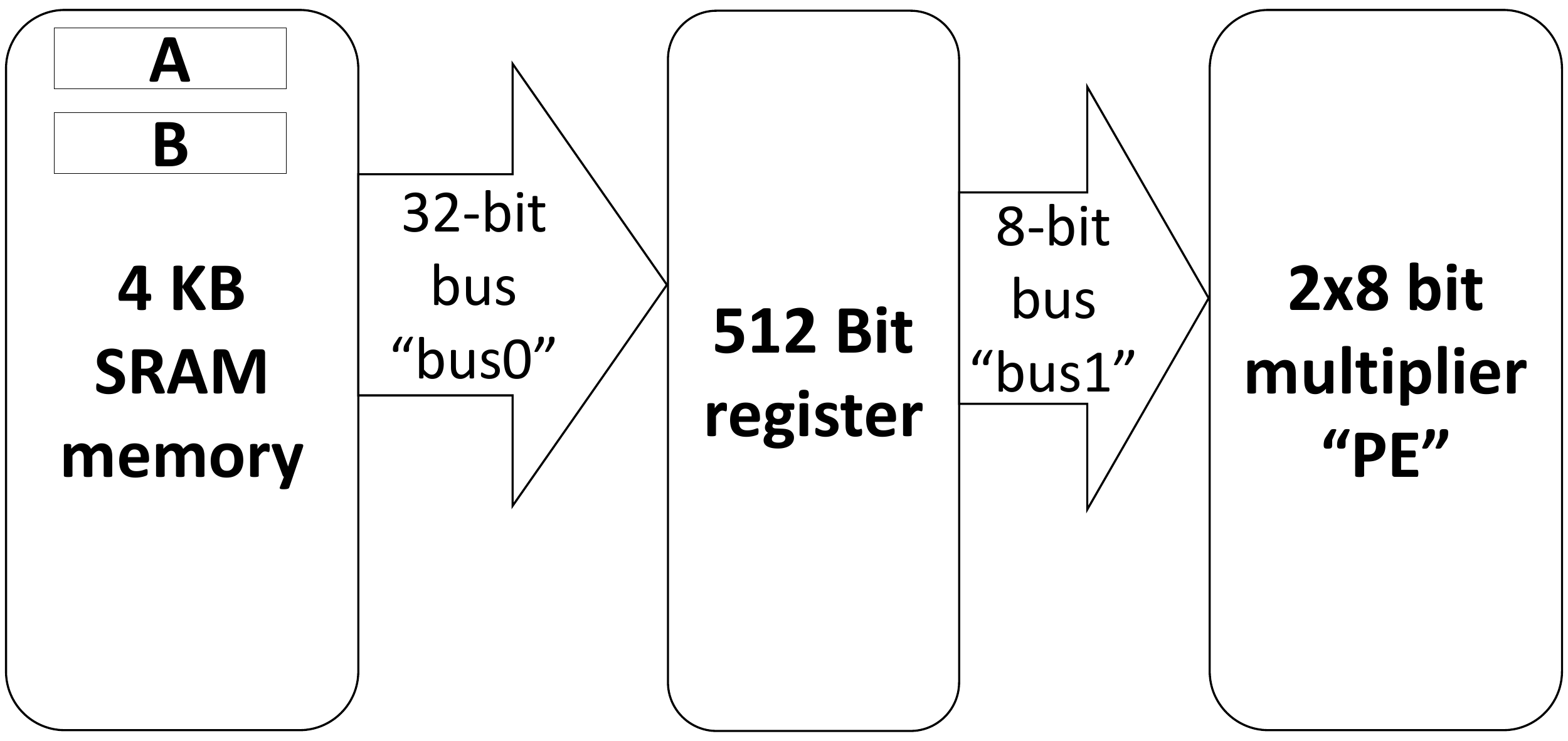}}
    \caption{Trivial memory and PE hardware layout}
	\label{fig:IntuitionExample}
\end{figure}

Before delving into the details of the DTT algorithm, we explain the
logic behind this method on a trivial example described in the Fig.
\ref{fig:IntuitionExample}. The key idea behind DTT is that a data
movement is the action of taking data in one place at some cycle and sending it
to another place, where it will arrive at an ulterior cycle. Both the spatial
(i.e. where the data is moving) and temporal (i.e. when the data is moving)
components must be taken into account.

\begin{figure*}[tbp]
\begin{subfigure}{.5\textwidth}
  \centering
  \includegraphics[width=.8\linewidth]{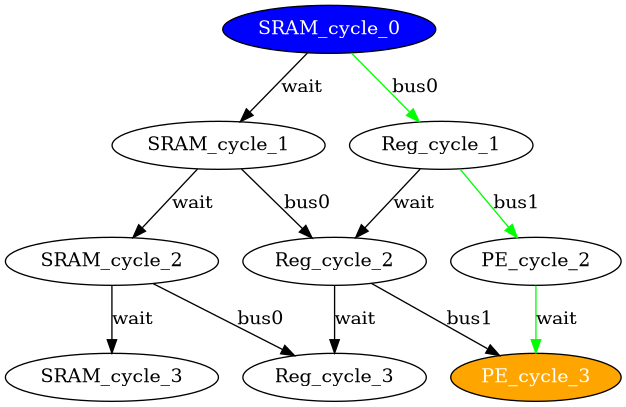}  
  \caption{Sending A to the PE}
  \label{fig:FirstExplanation_A}
\end{subfigure}
\begin{subfigure}{.5\textwidth}
  \centering
  \includegraphics[width=.8\linewidth]{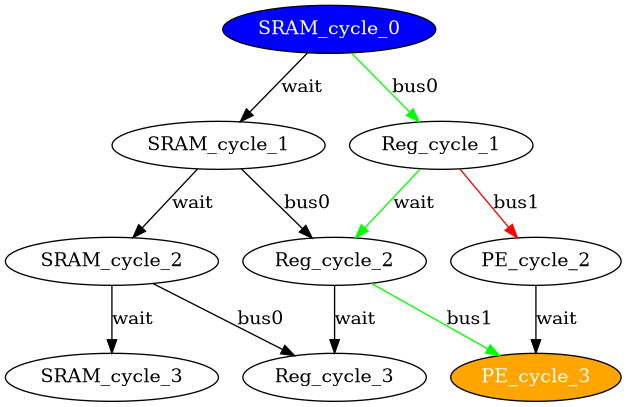}  
  \caption{Sending B to the PE}
  \label{fig:FirstExplanation_B}
\end{subfigure}
\caption{Scheduling graph for trivial example in Fig. \ref{fig:IntuitionExample}}
\label{fig:FirstExplanation_both}
\end{figure*}

Let us consider in Fig. \ref{fig:IntuitionExample} that we are trying to perform
a multiplication between two 8 bits operands A and B initially loaded in the
SRAM at cycle 0. The interconnect between SRAM and register is 32 bits wide
while the one between register and multiplier is 8 bits wide. Both interconnect
buses have a latency of 1 clock cycle.
The proposed DTT algorithm would reach its solution in two steps.
First, it notices that the only way to perform the multiplication is to send the
operands A and B to the multiplier and that the first cycle where the required
data can actually reach the PE is the cycle 3 (because of the bottleneck of
``bus1"). We shall refer to this step in DTT as the \textbf{placing step}.
Next, we use Dijkstra's algorithm to find how to send both A and B
to the PE at cycle 3. To do this, we are going to find a path on an abstract
scheduling graph, shown in Fig. \ref{fig:FirstExplanation_both}. 
In short, we are using Dijkstra's algorithm to explore the different
schedules and choose the best one where A and B reach the PE at cycle 3. We
shall refer to this step as the \textbf{routing step} of the DTT method.

Dijkstra's algorithm must route the operands one by one. 
The scheduling graph for the data movement of A is shown on the
Fig. \ref{fig:FirstExplanation_A}. The source node is shown in blue, the
target node in orange, and the solution of the algorithm in green. Fig.
\ref{fig:FirstExplanation_B} shows the scheduling graph when moving operand B.
The link in red indicates that ``bus1" is unavailable between the cycle 1 and 2,
because it is already carrying the operand A which occupies all the available
bandwidth (i.e. 8 bits).
In simple terms, the data movement schedule presented in Fig.
\ref{fig:FirstExplanation_A} is
\begin{enumerate}
	\item First move A from SRAM to the register at cycle 0.
	\item Then move A from the register to the PE at cycle 1.
	\item Wait inside the PE until cycle 3.
\end{enumerate}
Similarly, for Fig. \ref{fig:FirstExplanation_B}, the movement of B is
\begin{enumerate}
	\item First move B from SRAM to the register at cycle 0.
	\item Then let B wait one cycle inside the register (as the 8 bit bus is busy moving A).
	\item Finally, move B from the register at cycle 2 and reach the PE at cycle 3.
\end{enumerate}
\subsection{Data structures for the Algorithm} \label{section:DataStructure}
In this section, we briefly describe the data structures that were used to
implement the DTT algorithm in our proof of concept Python code.
The four main classes used for the implementation are the \textit{Datum} class,
the \textit{Node} class, the \textit{Wire} class and the \textit{Actor} class.

The \textit{Datum} class is used to represent a single unit of information
(eg. a byte) in the algorithm. For instance in Fig.
\ref{fig:IntuitionExample}, A and B would each be a distinct \textit{Datum}. In
terms of code implementation, a \textit{Datum} can be a simple integer.

The \textit{Node} class represents some memory element in the physical
hardware and is used to build the scheduling graph. In Fig.
\ref{fig:IntuitionExample} ``SRAM", ``Reg" and ``PE" would be instances of the
\textit{Node} class. The \textit{Node} class represents a node in the hardware
layout and not a node in the scheduling graph, which contains an additional
temporal information, as seen in Fig. \ref{fig:FirstExplanation_both}. The
history of the \textit{Node} is implemented with  a dictionary which yields for
any clock cycle the list of \textit{Datum} present in the \textit{Node}.

The \textit{Wire} class represents a path for data to move in the hardware
layout. In the Fig. \ref{fig:IntuitionExample}, ``bus0" and ``bus1" would be
\textit{Wire} instances. Note that not all \textit{Wire} instances refer to
physical wires. To let a \textit{Datum} wait inside a \textit{Node}, a pseudo waiting \textit{Wire} 
is drawn, which is simply a \textit{Wire} going from the \textit{Node} to itself.
The \textit{Node} and \textit{Wire} classes are used during the routing step of
the DTT algorithm. 

An \textit{Actor} class is a dedicated class used in during the placing step describing some 
hardware element which can perform a specific action. For instance in Fig. \ref{fig:IntuitionExample}, ``PE" is an
\textit{Actor} which is able to perform a multiplication. The \textit{Actor} class can
also represent more complex hardware as described in section \ref{section:PlacingStep}.

\subsection{Placing Algorithm}
\label{section:PlacingStep}
The purpose of the placing step of DTT is to assign the operations to
\textit{Actor} instances such that it shall lead the routing step to find the
best solution. But since the routing step is expensive, the placing step
must rely on a simpler, less granular representation of the state of the
hardware. This is analogous to the place and route algorithm used for FPGAs.
Unlike for the routing step, which is done using Dijkstra's algorithm, there is
no single algorithm which can be used for the placing step of DTT.
Heuristics can be used to obtain a better placement, which will lead to
a better overall scheduling.

In this section, we are going to describe a simple placing algorithm, the
\textit{First Best Fit} algorithm. One requirement for this algorithm is to have some \textit{affinity}
function, which scores how fit an \textit{Actor} is for some data. One possible
\textit{affinity} function would be to count the values in the input data
already cached by the \textit{Actor}. The high-level code for this
\textit{First Best Fit} placing is provided in Algorithm \ref{alg:FirstBestFit}.
\begin{algorithm}[tbp]
    \caption{First Best Fit placing algorithm}    \label{alg:FirstBestFit}
    \begin{algorithmic}
        \FORALL{operations to schedule}
            \IF{any actor is available}
                \STATE Take the Actor with highest affinity for the current operation.
                \STATE Assign the operation to this Actor.
                \IF{the Actor is full}
                    \STATE Mark the Actor busy for $X$ cycles.
                \ENDIF
            \ELSE
                \STATE Wait one cycle, the Actors might already be busy.
            \ENDIF
        \ENDFOR
    \end{algorithmic}
\end{algorithm}
The exact condition on which an \textit{Actor} becomes busy depends on the
context. In this paper we will consider reconfigurable MAC arrays as PEs,
which become busy when all their multipliers are in use for one or more dot
products.
The output of the placing step should be a succession of instructions holding at
least 3 pieces of information: \textbf{which} list of \textit{Datum} should be
sent, \textbf{where} they should go (i.e. which \textit{Actor}) and
\textbf{when} they should arrive.
\subsection{Routing Algorithm}
The core step of DTT is the routing algorithm, performed with Dijkstra's
algorithm applied on the scheduling graph. The output of the placing step
gives information about what data should move, and the routing step finds how or through what 
to optimally move the data. The high-level pseudo-code for this routing step is
given in Algorithm \ref{alg:RoutingStep}.
\begin{algorithm}[tbp]
    \caption{High-level pseudo-code of the routing step}
    \label{alg:RoutingStep}
    \begin{algorithmic}
        \FORALL{operations to route}
            \STATE Get the target of the path (i.e. the \textit{Node} + cycle where the data is needed).
            \FORALL{Datum in the input of the operation}
                \STATE Find all the locations (\textit{Node} + cycle) where the \textit{Datum} is present throughout the history.
                \STATE Find path between any of the sources and the target.
                \FORALL{\textit{Node} and cycle in the path}
                    \STATE Mark in the \textit{Node} that at the given cycle, it holds the involved \textit{Datum}.
                \ENDFOR
                \STATE Mark the Wires involved in a similar manner.
            \ENDFOR
        \ENDFOR
    \end{algorithmic}
\end{algorithm}
A part of this routing step is illustrated on the simple example of the Fig. 
\ref{fig:FirstExplanation_both}. We can see that the operand A goes from ``SRAM"
at cycle 0 to ``Reg" at cycle 1, and from there reaches ``PE" at cycle 2, where it
remains until cycle 3. As the algorithm \ref{alg:RoutingStep} states, we would
write this information in the associated \textit{Node} and \textit{Wire}
instances. Thus, we can deduce that the operand B cannot go through ``bus1" at
cycle 1 (recall ``bus 1" is busy moving A, Fig. \ref{fig:FirstExplanation_B}).
\section{Case Studies}
\subsection{Step by Step Example}
\begin{figure}[tbp]
    \scalebox{0.9}{\includegraphics[width=\linewidth]{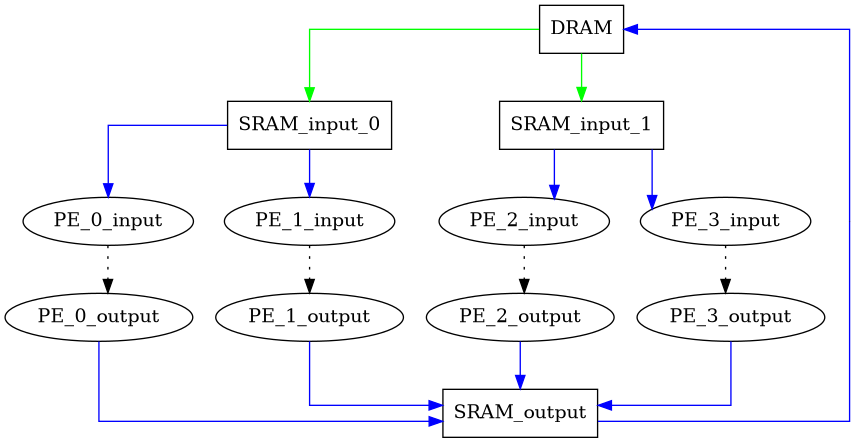}}
    \caption{Simple hardware used for the detailed example}
    \label{fig:Example1}
\end{figure}
In this section, we are going to do a step by step run of DTT on a
simple reproducible example, which can be used to test a DTT
implementation. The hardware we are considering is described in Fig.
\ref{fig:Example1}. It contains a global DRAM, two input SRAM, four processing
elements and one output SRAM. All the sizes, costs and delays of the elements of
the hardware are given in the Table \ref{tab:TableExample1}.
\begin{table}[tbp]
    \caption{Detailed information about the hardware of Fig. \ref{fig:Example1}}
	\label{tab:TableExample1}
	\centering
    \begin{tabular}{| l || c | c | c |}
        \hline
        element\_name & size & cost & delay \\ \hline
        \hline
        DRAM & 512B & $\emptyset$ & $\emptyset$ \\ \hline
        SRAM\_* & 64B & $\emptyset$ & $\emptyset$ \\ \hline
        PE\_*\_input & 4B & $\emptyset$ & $\emptyset$ \\ \hline
        PE\_*\_output & 2B & $\emptyset$ & $\emptyset$ \\ \hline
        green bus & 16B & 10 & 4 \\ \hline
        blue bus & 1B & 2 & 1 \\ \hline
        wait wire & $\infty$ & $\varepsilon$ & 1 \\ \hline
	\end{tabular}
\end{table}
The information in Table \ref{tab:TableExample1} is enough to build the
scheduling graph for the routing step. The PEs also must be described for the
placing step, with a simple black box model provided in Table
\ref{tab:TableExample1Placing}. Some of this information was estimated, and
there might be more than one way to determine the right values to use. 
For instance, the Actor cool down, i.e. the delay in cycles that should be left
between two sucessive operations, is usually determined through trial and error,
as an underestimated value would lead to a timing fault in the routing step.

\begin{table}[tbp]
    \caption{Detailed information for the placing step of Fig. \ref{fig:Example1}}
	\label{tab:TableExample1Placing}
	\centering
    \begin{tabular}{| c | c | c |}
		\hline
		Field & Value & Comment \\ \hline \hline
        Actor cooldown $X$ & 1 cycle & see algorithm \ref{alg:FirstBestFit}\\ \hline
        multiplication delay & 1 cycle & after which the result is ready \\ \hline
        distribution latency & 8 cycles & estimated time to fetch the operands \\ \hline
        memory size & 4 bytes & size of the operand buffer \\ \hline
        multiplier count & 2 & inferred from the memory size \\ \hline
    \end{tabular}
\end{table}
Before any computation is scheduled, the initial memory content must be
described to DTT in order for it to know where each \textit{Datum} can be
found. This would be done by arbitrarily adding the \textit{Datum} to the
history of the DRAM \textit{Node} (see the section \ref{section:DataStructure}).
In this example, we initially load the data numbered $0$ to $5$ in the DRAM.
The workload used for the example is described in Table
\ref{tab:TableExampleWorkload1}, with an offset on the last operation to handle
data dependency. Note that we consider dot products for our operations, that is because of the operations which can be sped up in ML computations involve some dot products. Examples of such operations include fully connected layers or convolutional neural networks.
\begin{table}[tbp]
    \caption{Example workload description and its placing result for example of Fig. \ref{fig:Example1}}
\begin{subtable}{.5\linewidth}
    \caption{Workload }
	\label{tab:TableExampleWorkload1}
	\centering
    \begin{tabular}{| l | c | c |}
        \hline
        Operation & Result & Offset \\ \hline
        \hline
        \shortstack{$0 \times 1$ \\ + $2 \times 3$} & $100$ & $0$ \\ \hline
        \shortstack{$1 \times 2$ \\+ $3 \times 4$} & $101$ & $0$ \\ \hline
        $2 \times 3$ & $102$ & $0$ \\ \hline
        $4 \times 5$ & $103$ & $0$ \\ \hline
        \shortstack{$100 \times 101 $ \\ \\+ $102 \times 103$ }& $104$ & $20$ \\ \hline
	\end{tabular}
  \end{subtable}%
  \begin{subtable}{.5\linewidth}
    \caption{Result of placing step}
	\label{tab:TablePlacing1}
    \centering
    \begin{tabular}{| l | c | c |}
        \hline
        Cycle & \textit{Node} & Data \\ \hline
        \hline
        8 & PE\_0\_input & \shortstack{[0, 1,\\ 2, 3]} \\ \hline
        8 & PE\_1\_input & \shortstack{[1, 2,\\ 3, 4]} \\ \hline
        8 & PE\_2\_input & [2, 3] \\ \hline
        8 & PE\_2\_input & [4, 5] \\ \hline
        28 & PE\_0\_input & \shortstack{[100, 101,\\ 102, 103]} \\ \hline
	\end{tabular}
  \end{subtable} 
\end{table}
Running the placing step for the provided workload yields the Table
\ref{tab:TablePlacing1}. The PEs are assumed to be reconfigurable, hence
several different dot products are scheduled together when multipliers are
available.
Before running the routing step, one must explain to DTT how the output values
$100$ to $104$ will appear in the hardware, here by arbitrarily adding a \textit{Datum}
in the history of the appropriate \textit{Node} instances (see the section
\ref{section:DataStructure}). This is similar to how \textit{Datum} are loaded
in the DRAM during the initialization. Once that has been taken care of, the
routing step can be pursued, and its result is provided in Table
\ref{tab:TableRouting1}.
\begin{table}[tbp]
    \caption{Detailed result of the routing step for Fig. \ref{fig:Example1}}
	\label{tab:TableRouting1}
	\centering
    \begin{tabular}{| l | c | c | c | c |}
        \hline
        \textit{Datum} & source \textit{Node} & start cycle & end \textit{Node} & end cycle \\ \hline
        \hline
        0 & DRAM & 3 & SRAM\_input\_0 & 7 \\ \hline
        0 & SRAM\_input\_0 & 7 & PE\_0\_input & 8 \\ \hline
        \textcolor{red}{1} & \textcolor{red}{DRAM} & \textcolor{red}{2} & \textcolor{red}{SRAM\_input\_0} & \textcolor{red}{6} \\ \hline
        \textcolor{red}{1} & \textcolor{red}{SRAM\_input\_0} & \textcolor{red}{6} & \textcolor{red}{PE\_0\_input} & \textcolor{red}{7} \\ \hline
        2 & DRAM & 1 & SRAM\_input\_0 & 5 \\ \hline
        2 & SRAM\_input\_0 & 5 & PE\_0\_input & 6 \\ \hline
        3 & DRAM & 0 & SRAM\_input\_0 & 4 \\ \hline
        3 & SRAM\_input\_0 & 4 & PE\_0\_input & 5 \\ \hline
        \textcolor{red}{1} & \textcolor{red}{SRAM\_input\_0} & \textcolor{red}{7} & \textcolor{red}{PE\_1\_input} & \textcolor{red}{8} \\ \hline
        2 & SRAM\_input\_0 & 6 & PE\_1\_input & 7 \\ \hline
        3 & SRAM\_input\_0 & 5 & PE\_1\_input & 6 \\ \hline
        4 & DRAM & 0 & SRAM\_input\_0 & 4 \\ \hline
        4 & SRAM\_input\_0 & 4 & PE\_0\_input & 5 \\ \hline
        2 & DRAM & 3 & SRAM\_input\_1 & 7 \\ \hline
        2 & SRAM\_input\_1 & 7 & PE\_2\_input & 8 \\ \hline
        3 & DRAM & 2 & SRAM\_input\_1 & 6 \\ \hline
        3 & SRAM\_input\_1 & 6 & PE\_2\_input & 7 \\ \hline
        4 & DRAM & 1 & SRAM\_input\_1 & 5 \\ \hline
        4 & SRAM\_input\_1 & 5 & PE\_2\_input & 6 \\ \hline
        5 & DRAM & 0 & SRAM\_input\_1 & 4 \\ \hline
        5 & SRAM\_input\_1 & 4 & PE\_2\_input & 5 \\ \hline
        \textcolor{blue}{100} & \textcolor{blue}{PE\_0\_output} & \textcolor{blue}{21} & \textcolor{blue}{SRAM\_output} & \textcolor{blue}{22} \\ \hline
        \textcolor{blue}{100} & \textcolor{blue}{SRAM\_output} & \textcolor{blue}{22} & \textcolor{blue}{DRAM} & \textcolor{blue}{23} \\ \hline
        \textcolor{blue}{100} & \textcolor{blue}{DRAM} & \textcolor{blue}{23} & \textcolor{blue}{SRAM\_input\_0} & \textcolor{blue}{27} \\ \hline
        \textcolor{blue}{100} & \textcolor{blue}{SRAM\_input\_0} & \textcolor{blue}{27} & \textcolor{blue}{PE\_0\_input} & \textcolor{blue}{28} \\ \hline
        \textcolor{blue}{101} & \textcolor{blue}{PE\_1\_output} & \textcolor{blue}{20} & \textcolor{blue}{SRAM\_output} & \textcolor{blue}{21} \\ \hline
        \textcolor{blue}{101} & \textcolor{blue}{SRAM\_output} & \textcolor{blue}{21} & \textcolor{blue}{DRAM} & \textcolor{blue}{22} \\ \hline
        \textcolor{blue}{101} & \textcolor{blue}{DRAM} & \textcolor{blue}{22} & \textcolor{blue}{SRAM\_input\_0} & \textcolor{blue}{26} \\ \hline
        \textcolor{blue}{101} & \textcolor{blue}{SRAM\_input\_0} & \textcolor{blue}{26} & \textcolor{blue}{PE\_0\_input} & \textcolor{blue}{27} \\ \hline
        \textcolor{blue}{102} & \textcolor{blue}{PE\_2\_output} & \textcolor{blue}{19} & \textcolor{blue}{SRAM\_output} & \textcolor{blue}{20} \\ \hline
        \textcolor{blue}{102} & \textcolor{blue}{SRAM\_output} & \textcolor{blue}{20} & \textcolor{blue}{DRAM} & \textcolor{blue}{21} \\ \hline
        \textcolor{blue}{102} & \textcolor{blue}{DRAM} & \textcolor{blue}{21} & \textcolor{blue}{SRAM\_input\_0} & \textcolor{blue}{25} \\ \hline
        \textcolor{blue}{102} & \textcolor{blue}{SRAM\_input\_0} & \textcolor{blue}{25} & \textcolor{blue}{PE\_0\_input} & \textcolor{blue}{26} \\ \hline
        \textcolor{blue}{103} & \textcolor{blue}{PE\_2\_output} & \textcolor{blue}{18} & \textcolor{blue}{SRAM\_output} & \textcolor{blue}{19} \\ \hline
        \textcolor{blue}{103} & \textcolor{blue}{SRAM\_output} & \textcolor{blue}{19} & \textcolor{blue}{DRAM} & \textcolor{blue}{20} \\ \hline
        \textcolor{blue}{103} & \textcolor{blue}{DRAM} & \textcolor{blue}{20} & \textcolor{blue}{SRAM\_input\_0} & \textcolor{blue}{24} \\ \hline
        \textcolor{blue}{103} & \textcolor{blue}{SRAM\_input\_0} & \textcolor{blue}{24} & \textcolor{blue}{PE\_0\_input} & \textcolor{blue}{25} \\ \hline
	\end{tabular}
\end{table}
A deeper look into the Table \ref{tab:TableRouting1} reveals two important
strengths of the DTT algorithm.  First, DTT will spontaneously optimize the data
movements for the operations. For instance, the \textcolor{red}{red part of the
Table \ref{tab:TableRouting1}} shows that the \textit{Datum} 1 was originally
sent to the PE\_0 for the first dot product, and DTT will fetch it from cache
instead of DRAM for the second operation. Similarly, notice \textcolor{blue}{in
blue} that DTT will naturally use the \textit{Datum} $100$ to $103$ to perform
the last dot product. DTT treats these values in the same way it treats values
coming from the DRAM. If neural network layers are represented as a sequence of dot products, DTT
will thus be able to optimize data movement across more than one layer at a
time unlike classic for-loop refactoring algorithms. Thus if weights are reused in a NN model, DTT will try to reuse them on chip as far as possible without reloading them from main memory. To enable this, the reused weights must be marked during workload description to DTT.
\subsection{Reduction Network}

\begin{figure}[tbp]
    \includegraphics[width=\linewidth]{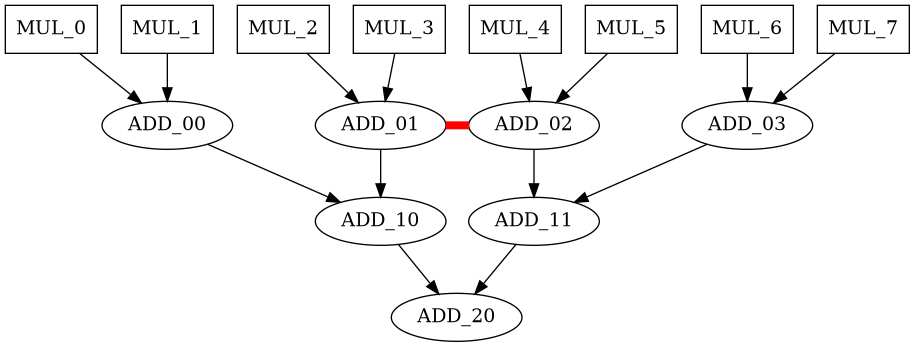}
    \caption{Example of using DTT for a reduction network}
	\label{fig:Example2}
\end{figure}

A reduction network is a
set of interconnected adders which will accumulate the result of some
multipliers. This is used to compute one or several dot products at the same
time. Scheduling reductions with DTT is not immediately obvious, as the
algorithm is naturally more suited for distributing values. In practice, one
only needs to pretend to distribute the results to the multipliers and play the
computed data flow backwards to schedule a reduction. This is because the
addition and data reuse operation have an opposite data flow.

\begin{figure}[tbp]
	\includegraphics[width=\linewidth]{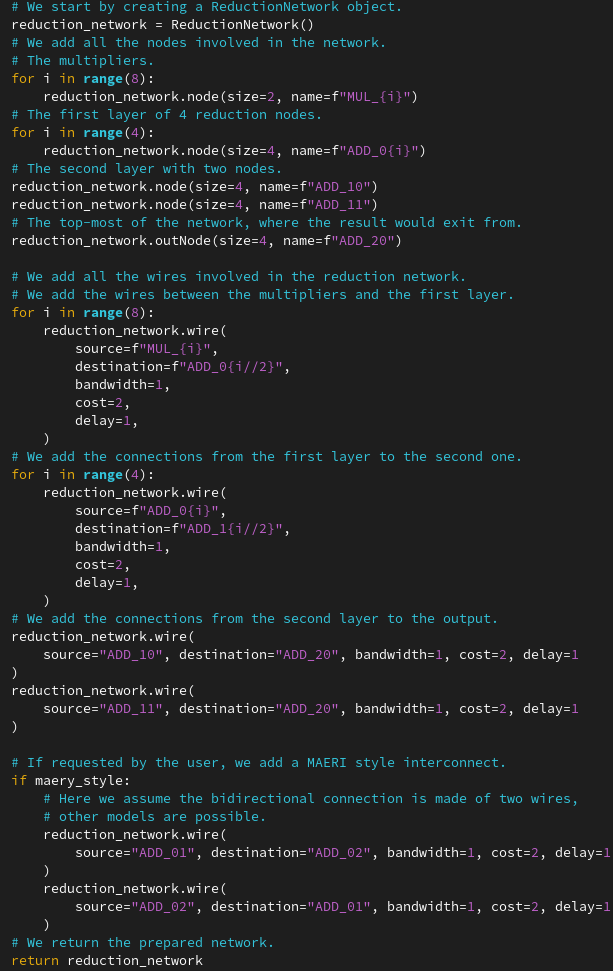}
	\caption{DTT code for the reduction network of Fig. \ref{fig:Example2}}
	\label{fig:ReductionCode}
\end{figure}

Most reduction networks take the form of a binary tree where the leaves are
multipliers and the nodes are adders. One limitation of such a reduction
network is that performing more than one dot product at a time on the network
will lead to a lot of blocking. To improve on this design, the paper \cite{maeri}
proposed the ``augmented reduction tree" recreated in Fig.
\ref{fig:Example2}, which is a binary tree with an additional connection between
ADD\_01 and ADD\_02.
Hypothetically, a hardware designer would want to benchmark the accelerator with
and without the ``augmented connection" to see if its benefits over weigh its
costs. DTT can be used to create a prototyping back end for this purpose, which
can be simply modified to compare the two reduction trees: simply remove from 
the description of the hardware graph the \textit{Wire} between the \textit{Node} 
instances ADD\_01 and ADD\_02 to benchmark the simple reduction tree, as shown in
Fig. \ref{fig:ReductionCode}.

Some information that designers might find interesting and can be easily extracted from DTT are:
\begin{enumerate}
	\item The \textbf{energy savings} for a given workload with the augmented
	connection,
	\item The \textbf{runtime difference} between the normal and augmented
	reduction trees,
	\item The \textbf{utilization rate} of the augmented connection in the
	workload,
\end{enumerate}
These low-level statistics would then be used to decide whether an augmented
reduction tree or a simple reduction tree should be included in the designed
hardware.
\subsection{Working with complex Hardware}
\label{section:ComplexHardware}
\begin{figure}[tbp]
    \includegraphics[width=\linewidth]{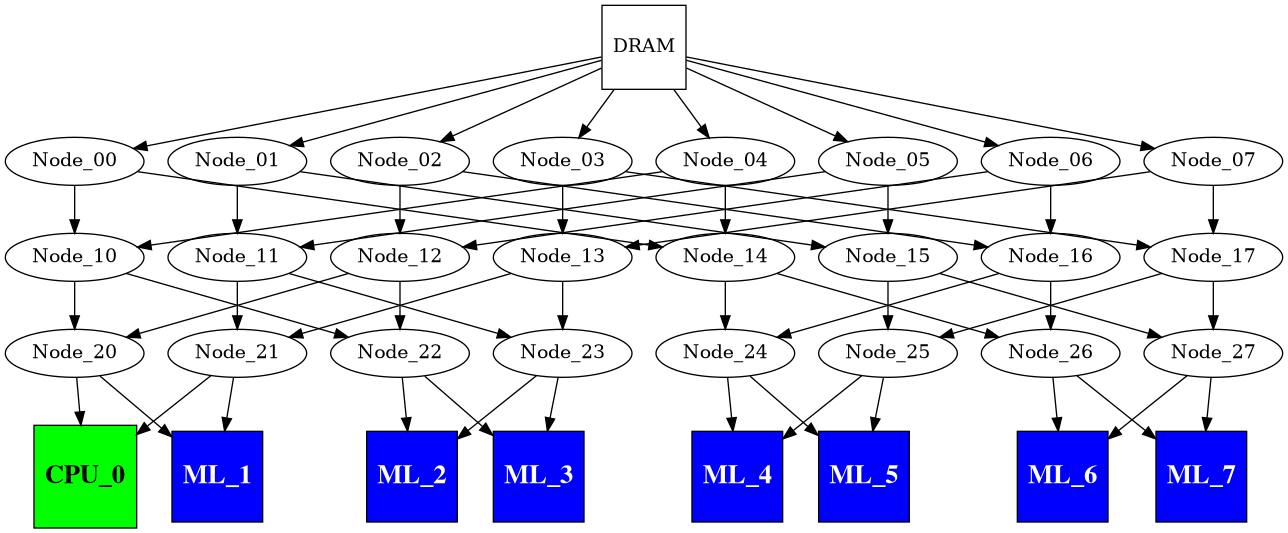}
    \caption{Example of complex hardware covered by DTT}
	\label{fig:Example3}
\end{figure}

For the last example, we are going to explain how a more complex hardware can be
handled by DTT. This is meant to show the flexibility of the algorithm in more
interesting cases.
The hardware for this example is shown in Fig. \ref{fig:Example3}. It uses a
butterfly-style distribution network to send data from a global DRAM to several
high-level components. In this design, seven ML accelerators are used (in \textcolor{blue}{blue} 
along with a general-purpose processor (in \textcolor{green}{green}
on the Fig \ref{fig:Example3}).
The idea here is that the NN model is made of several types of operations. Let
us assume that only dot products, matrix multiplications and custom functions
are used here. The dot products and matrix multiplications can be carried out by
either the CPU or an accelerator, but an accelerator would be faster and more
efficient. The rare custom functions however may only be performed out by the
CPU.
We are going to modify the placing step described in Algorithm \ref{alg:FirstBestFit}. Since not every PE can carry out every operations, we first have to filter for the provided operation code the \textbf{Actors} which are relevant for the computation. The new placing step is presented in \ref{alg:ModifiedPlacing}.
\begin{algorithm}[tbp]
    \caption{Placing Algorithm tuned for Fig. \ref{fig:Example3}}
    \label{alg:ModifiedPlacing}
    \begin{algorithmic}
        \FORALL{operations to schedule}
	    \STATE Filter the \textit{Actors} which can perform the operation
	    \STATE Use First Best Fit to place the operation among those valid \textit{Actors}
        \ENDFOR
    \end{algorithmic}
\end{algorithm}
The routing step however would remain unchanged. This may come as a surprise,
since several different operations are involved, but as mentioned before the
only information considered for the routing step are: \textbf{what} data should
be moved, \textbf{where} is it going to and \textbf{when} should it arrive.
For as far as the routing algorithm is concerned, moving data for a dot product
is no different than moving it for a custom computation. This also implies that
DTT will spontaneously optimize data movements across different operations.

\begin{figure}[htbp]
    \includegraphics[width=\linewidth]{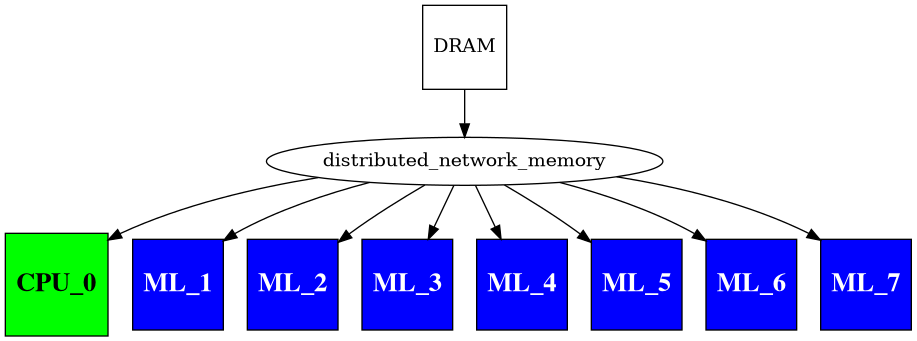}
    \caption{Simpler description of Fig. \ref{fig:Example3} for higher level tools}
	\label{fig:Example3_bis}
\end{figure}

When benchmarking such hardware, a new issue might arise: the upstream
DSE tool used probably does not understand what a butterfly network is. One
would provide a simpler hardware description to optimize the workload for, and
then take the optimized neural network and use DTT to schedule it for the
complex hardware. For instance, the hardware of Fig. \ref{fig:Example3} could be
described as Fig. \ref{fig:Example3_bis} to a tool such as ZigZag or Timeloop.
Although giving an inaccurate description of the accelerator to the
DSE tool could mean that its solution will be sub-optimal, it should
still give a much better starting point for DTT.

\section{DTT-based NoC}
In this last part, we are going to present and end to end theoretical architecture that would take advantage of the DTT algorithm to optimize data movements in its NoC. In this architecture, the \textit{Datum} considered will thus be entire memory pages. 

\subsection{Compiler overview}

The compiler for the architecture would take as input a single threaded stream of operations to execute on the hardware. Its first task is to multi-thread those tasks between the various PEs in the architecture by running the placing step of the DTT algorithm. This approach is reminiscent of VLIW compilers\cite{vliw_compiler}, except that VLIW compilers also have to deal with runtime branches, while those do not appear in DTT’s deterministic workloads. As explained in section \ref{section:ComplexHardware}, this placing step can accommodate heterogeneous PEs.

Once the placing step of DTT is over, the compiler knows when each operation should be performed and where they should take place. The routing step is used to decide how to send the data to the PEs so that they will be able to start their computations on time. The order in which data movement will happen in the NoC is:

\begin{enumerate}
	\item The controller asks a source \textit{Node} to send a page to another PE
	\item The data movement takes place and the PE receives the page
	\item The controller sends the instruction to the PE and the computation starts
\end{enumerate}

Because data arrives before the PEs start their operations, the PEs will never cache miss. In practice though, DTT has two weaknesses which prevent us from using it directly: 

\begin{itemize}
	\item The routing step is too expensive to be performed on an edge device, 
	\item The output of the routing step (a list of data movements) is too big to be reasonably sent to an edge device. 
\end{itemize}

Fortunately, there is a known solution to those issues, which is inspired by macro-routing\cite{macro_routing}. The idea is to provide DTT with checkpoints that it can use to forget what came before. Hence, once the compiler is done performing the routing step, it only saves way-points (i.e. a few specific points the path has to traverse in order) from the paths instead of the whole path. Those way-points will be placed by the compiler into a binary file along with the scheduled operations for the accelerator. 

\subsection{DTT controller}

Because we rely on DTT, the PEs used in our theoretical architecture do not use a distributed cache coherence protocol, but instead a global hardware controller connected to the NoC. This controller is here intended to be a small CPU core, although dedicated hardware might also be used. This controller has three main responsibilities:

\begin{enumerate}
	\item It reads the input program for the accelerator
	\item It reconstructs the paths for the data movements from the way-points
	\item It sends control packets to the PEs in order to have them realize the data movements it computed
\end{enumerate}

The way points computed by the compiler and saved into the program file are sent to the hardware controller, which will run Dijkstra’s algorithm again to fill in the missing movements between the way-points at run-time. This technique could be seen as analogous to compressing a video to only its key-frames, then filling the intermediary frames via interpolation\cite{video_compression}. After the reconstruction of paths from the DTT algorithm, each data movement has to be turned into a control packet that will be sent by the controller to the relevant PEs. 

The control packets would be sent over the NoC using dynamic routing protocols (as opposed to a fixed route decided by the controller) and should have plenty of time to reach their targets before the data movement needs to be carried out. To be sure that the control packets are consumed at the right time, a simple TTL (Time-toLive) mechanism could be used. Once the control packets are consumed, they will trigger a 2-way DMA access, sending the data from the source to the target (without going through the controller) and following the route decided by DTT. This ensures no unexpected congestions will appear in the NoC, as all the data movements have been foreseen by the controller. 

\subsection{Comparison with directory-based protocol}
 
\begin{figure}
	\begin{subfigure}{.5\linewidth}
		\centering
		\includegraphics[width=.9\linewidth]{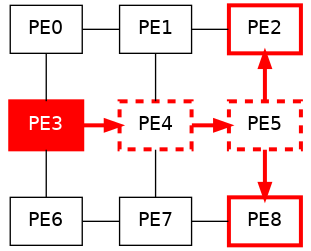}
		\caption{Sharing a single data transfer between two targets}
		\label{fig:first_movement}
	\end{subfigure}%
	\begin{subfigure}{.5\linewidth}
		\centering
		\includegraphics[width=.9\linewidth]{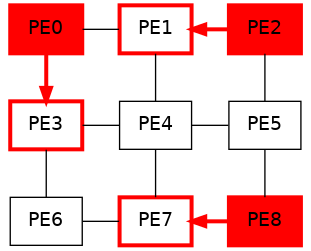}
		\caption{Fetching page multiple copy of the same page from several sources}
		\label{fig:second_movement}
	\end{subfigure}
\caption{Illustration of possible data-movement optimizations in DTT}
\label{fig:data_movement_optimizations}
\end{figure}

One significant asset of using DTT, compared to other coherent networking approaches, is its ability to optimize the data movements. Two possible traffic optimizations with DTT are shown in Fig. \ref{fig:data_movement_optimizations}. During the routing step, DTT will use Dijkstra's algorithm to find the shortest way to bring the required data to the PEs for the operations, leading to an expected improvement in the overall performance of the accelerator. This works because the cache coherence is ensured by one omniscient actor, hence finding correct optimizations becomes much easier. 

The controller used in the theoretical architecture creates a bottleneck for the execution, as every operation needs to be processed by the controller before being executed in the chosen PE. Thus, a DTT based architecture is expected to not scale as well as a directory-based architecture, which is summarized in Table \ref{tab:TableCoherence}.

\begin{table}
    \caption{Comparison of DTT with classic coherence protocols}
	\label{tab:TableCoherence}
	\centering
    \begin{tabular}{| l || c | c | c |}
        \hline
        Protocol & Bus-snooping & DTT & Full-directory \\ \hline
        \hline
	    Bottleneck & Bus bandwidth & Controller compute & Node distance \\ \hline
	    Best for & Small NoC & Medium NoC & Large NoC \\ \hline
	    Determinism & No & Required & No \\ \hline
	\end{tabular}
\end{table}

\section{Conclusion}
In this paper we presented DTT, a low complexity ahead of time hardware
scheduling method. We explained how the algorithm works and detailed important
parts of its implementation. Several examples were shown to show how DTT can
handle reconfigurable and custom hardware. Finally, we highlighted that DTT
opens new opportunities for hardware designers by enabling them to quickly draft
prototyping back ends. As mentioned at the beginning of the paper, DTT is
currently a proof of concept. As future improvement we would like to show its
usage with an application on a real ML workload.

\bibliographystyle{IEEEtran}
\bibliography{paper} 

\end{document}